\documentclass[apj]{emulateapj}
\usepackage{epsfig}
\usepackage{subfigure}

\def\Kepler{\textit{Kepler}}

\begin{document}
\title{An Efficient Automated Validation Procedure for Exoplanet Transit Candidates}

\author{Timothy D. Morton\altaffilmark{1}}
\email{tdm@astro.caltech.edu}
\altaffiltext{1}{Department of Astrophysics,
  California Institute of Technology, MC 249-17, Pasadena, CA 91125}

\begin{abstract}
Surveys searching for transiting exoplanets have found many more candidates than they have been able to confirm as true planets.  This situation is especially acute with the {\it Kepler} survey, which has found over 2300 candidates but has confirmed only 77 planets to date.  I present here a general procedure that can quickly be applied to any planet candidate to calculate its false positive probability.  This procedure takes into account the period, depth, duration, and shape of the signal; the colors of the target star; arbitrary spectroscopic or imaging follow-up observations; and informed assumptions about the populations and distributions of field stars and multiple-star properties.  Applying these methods to a sample of known {\it Kepler} planets, I demonstrate that many signals can be validated with very limited follow-up observations: in most cases with only a spectrum and an AO image.  Additionally, I demonstrate that this procedure can reliably identify  false positive signals.  Because of the computational efficiency of this analysis, it is feasible to apply it to all {\it Kepler} planet candidates in the near future, and it will streamline the follow-up efforts for {\it Kepler} and other current and future transit surveys.
\end{abstract}

\section{Introduction}

The first exoplanets found to transit the face of their host stars were all initially discovered by Doppler surveys, which detected minute radial velocity (RV) variations of stars indicative of the gravitational tug of orbiting planets \citep{charbonneau2000,henry2000,butler2004,bouchy2005,sato2005}.  Surveys designed to {\it detect} exoplanets via their transits soon followed \citep{alonso2004,mccullough2006,bakos2007,cameron2007}, motivated by the wealth of physical information that the light curve of a transiting planet can provide \citep{seager2003}.  While initial forecasts suggested that such programs would churn out planet detections at tremendous rates \citep{horne2003}, the actual yield of ground-based surveys has been much more modest, with only about a dozen discoveries published in the first five years of transit survey operations.  

One of the major reasons for this slower-than-hoped-for discovery rate (in addition to the various contributing subtleties discussed in \citet{beatty2008}) is the preponderance of various eclipsing stellar binary scenarios mimicking planet transit signals, or so-called ``astrophysical false positives,'' combined with the quantity and expense of the observational resources required to confirm a signal as truly planetary.    The classic example of an astrophysical false positive is the case of OGLE-TR-33 \citep{torres2004}.  Identified by the Optical Gravitational Lensing Experiment (OGLE) as a transiting planet candidate with a periodic photometric dip of 2\%, subsequent follow-up observations seemed only to confirm its planetary nature:  periodic RV variations were discovered in phase with the transit signal, no nearby stars were observed to be blended within the photometric aperture, and even the eclipse depth was measured to be color-independent.  However, further investigation revealed that various blended eclipsing binary scenarios could fit the light curve just as well as a transiting planet model, and upon careful inspection of the spectral time-series data it became clear that OGLE-TR-33 was indeed a hierarchical triple system containing an eclipsing binary, and not a transiting planet at all---the apparent Doppler shift of the lines was actually an asymmetry in the combined line profiles of two stars.  There were early isolated attempts to statistically quantify the probability that signals might be false positives \citep[e.g.]{sahu2006}, but this and other similar examples, in addition to theoretical predictions \citep{brown2003}, were strong early warnings to the community that false positive scenarios may lurk behind any transit candidate, and that extremely careful analysis is required to rule out false positives in any individual case.  

As a result, the traditional method of promoting a transit candidate signal to the status of a {\it bona fide} planet involves an in-depth, multi-instrument follow-up campaign. A prototypical example of this procedure is detailed in the discovery paper of TrES-1b \citep{alonso2004}, which was the first transiting exoplanet to be discovered by a survey specifically designed for the purpose.  The follow-up observations required to confirm this signal as a planet were the following:
\begin{itemize}
\item $H$- and $K$-band adaptive optics (AO) imaging to rule out contaminating stars of similar brightness outside of $\sim$$0\farcs3$ separation from the target star. 
\item Medium-resolution Echelle spectroscopy (velocity resolution of $8.5~{\rm km~s}^{-1}$) at seven epochs, to rule out scenarios involving massive binary companions and to characterize the host star as suitable for further follow-up.
\item Multi-color transit photometry with three different facilities in seven different filters (in addition to the discovery data), to check that the eclipse depth is color-independent, consistent with the eclipse of a cold, substellar object.
\item High-precision ($\sim$10 m s$^{-1}$) RV measurements with the Keck High Resolution Echelle Spectrometer (HIRES) at eight epochs, which detected the sinusoidal Doppler variation caused by the transiting planet.
\end{itemize} 
The ``funnel'' procedure established with this discovery, in which a large number of candidates\footnote{Tres-1b was one of 16 transit-like signals among 12,000 stars monitored by the survey in its field and the only one to become positively identified as a planet} are whittled down by various follow-up observations to identify the few that are worthy of observations with a large telescope such as Keck, was absolutely necessary in the early days of transit surveys, given the scarcity of precise RV spectrometers and the desire to discover and characterize individual exoplanetary systems.  Throughout the last decade this has been widely adopted as the follow-up paradigm: find many candidates, identify a subset appropriate for follow-up, and eventually confirm a small number of these with RV measurements.
 Understandably, most of the exoplanets detected in this manner have been hot Jupiters, given that both their photometric (detection) and Doppler (confirmation) signals are more readily measurable than those of smaller planets.

The {\it Kepler} mission \citep{borucki2008} poses completely different problems, both in scale and specifics, than do ground-based surveys.  Not only is the sheer volume of candidates completely unprecedented (over $2300$ \citep{batalha2012}), but the keystone of traditional confirmation (RV measurements) is literally impossible to obtain for most of the candidates, given current and immediately foreseeable telescope resources: the stars are too faint and/or the planets are too small.  Systems that show multiple transiting candidates sometimes display transit timing variations (TTVs) that may be used to dynamically confirm the planets in the system \citep{holman2010,lissauer2011,cochran2011,fabrycky2012,ford2012,steffen2012}, but the number of systems for which this is feasible is still an overwhelming minority of all the candidates.

Given these difficulties, it has become necessary to adopt a new paradigm of transiting planet confirmation: {\it probabilistic validation}.  The philosophy behind this strategy is to demonstrate that a particular transit candidate is much more likely to be a transiting planet than it is to be a false positive---which may be the only option available if positive confirmation via RVs or TTVs is impossible or impractical. This has been exemplified by the {\it Kepler} team's BLENDER procedure, which is descended from the methods which helped identify OGLE-TR-33 as a false positive, and also used to help confirm other OGLE candidates \citep[e.g.]{torres2005}.  Using a suite of follow-up spectroscopic and imaging observations combined with extensive light curve fitting, BLENDER rules out regions of parameter space where false positive models are unable to fit the photometric data as well as a planet model.  First demonstrated with the validation of Kepler-9d \citep{torres2011}, this procedure has been used to validate many of the planets which have so far received official {\it Kepler} number designations.  

However, despite the successes of BLENDER, {\it Kepler} still has a problem of scale: BLENDER is expensive, both in CPU- and person-hours, and has not yet been applied to large numbers of candidates in a wholesale manner.  As the overall mission of the {\it Kepler} project is not individual planet detections (as has been the case with ground-based surveys) but population statistics, there was  a need to estimate the {\it a priori} false positive probabilities (FPPs) for {\it Kepler} candidates, before any follow-up observation or detailed BLENDER-style analysis.  This was the inspiration for the work of \citet{mj11}, which demonstrated that for transit candidates that have passed the vetting tests that are possible with {\it Kepler} photometry alone, the FPP is typically around 5\% or less.\footnote{The dominant reason why this is much lower than has been the case for ground-based surveys is because the quality of the {\it Kepler} photometry allows for much more constraining pre-followup vetting than has been possible with ground-based surveys; in particular, {\it Kepler} is able to identify many blended binary scenarios by measuring the target's center-of-light to shift during eclipse.  This makes the effective ``blend aperture'' for {\it Kepler} much smaller than for ground-based wide-field surveys. Additionally, the precise photometry can put strict limits on the presence of any secondary eclipse, which further constrains false positive scenarios.  And finally, the signals are typically shallower than detected in ground-based surveys, which also contributes to intrinsically lower FP rates.}   For broad-brush studies to date, this result has been used to essentially ignore false positive contamination when analyzing the statistical properties of the candidate sample.  


Despite the reassurance these {\it a priori} calculations provide that the broad characteristics of the overall {\it Kepler} candidate sample reflect those of the true exoplanet population, there remains significant motivation to push {\it a priori} FPP analysis beyond MJ11.  First of all, MJ11 did not fully quantify all false positive scenarios; in particular, grazing eclipse signals,  non-blended eclipsing binaries, and background blended transiting planets were not considered.   A more thorough analysis will further increase the fidelity of the entire {\it Kepler} sample and thus enable more nuanced statistical analyses.   

Secondly, the framework presented in MJ11 uses only minimal information about the photometric signals; that is, their depth and period.  Much more information is available, particularly regarding the shape of the transit.  In addition, MJ11 did not provide a transparent way to incorporate the results of follow-up observations to update the {\it a priori} calculations.  And finally, as the process of probabilistically validating a planet is the same as demonstrating that its FPP is sufficiently low, a more stringent FPP calculation in the style of MJ11 may be used to validate large numbers of transit candidates for which traditional confirmation is difficult or impossible.  In addition to enabling ground-based surveys to streamline follow-up efforts, such a tool could ensure that the legacy of the {\it Kepler} mission is indeed thousands of transiting planets, not just planet candidates.  

This paper presents a comprehensive transit candidate validation procedure that is based on the {\it a priori} framework of MJ11, yet also uses information about the shape of the transit signal and naturally incorporates (yet does not require) follow-up spectroscopic and imaging observations.    An early version of this analysis, representing a halfway point between MJ11 and this work, was used to validate the KOI-961 planets \citep{muirhead2012}.  Crucially, this method is capable of analyzing large numbers of candidates quickly; the entire end-to-end computation for a single transit candidate signal takes only about 10 minutes on a typical personal computer.  It is thus capable of being applied to the entire {\it Kepler} candidate sample in the immediately foreseeable future.   Looking beyond {\it Kepler}, the larger goal of this procedure is to revolutionize the follow-up strategy of large-scale transit surveys by reducing the time and and cost necessary to identify false positives or confirm transiting planets, and to enable statistical analyses of transit candidate populations without having to wait for every candidate to be individually solved.

In \S\ref{sec:framework} I review the probabilistic framework introduced in MJ11 and summarize the false positive scenarios considered in this work.  I describe the entire procedure in detail in \S\ref{sec:procedure}, present the results of applying this procedure to a number of previously studied {\it Kepler} signals in \S\ref{sec:tests}, discuss the relationship of this work to MJ11 in \S\ref{sec:MJ11}, and provide concluding remarks in \S\ref{sec:discussion}.

\section{Framework}
\label{sec:framework}

Validating a transiting planet means demonstrating that the false positive probability (FPP) of the signal is small enough to be considered negligible.  Regardless of exactly where this threshold lies, the process of validation is the same as carefully calculating the FPP.  Following MJ11, I define the FPP for a given transit signal:

\begin{equation}
\label{eq:FPP}
{\rm FPP} = 1 - \Pr({\rm planet} ~|~{\rm signal})
\end{equation}
where the vertical line means `given' (i.e.~``the probability of there being a planet given the observed signal''), and
\begin{equation}
\label{eq:simplebayesLs}
\Pr({\rm planet} ~|~{\rm signal}) = \frac{\mathcal L_{\rm TP} \pi_{\rm TP} } { \mathcal L_{\rm TP} \pi_{\rm TP} +  \mathcal L_{\rm FP} \pi_{\rm FP} }.
\end{equation}

In the above equation, $\mathcal L$ represents the Bayesian likelihood factors (the ``probability of the data given the model''), and $\pi$ represents the priors (the {\it a priori} probabilities that each given scenario exists).  The TP subscript represents the true transiting planet scenario, or ``true positive,'' and FP represents all of the false positive scenarios.   

The false positive scenarios I consider in this analysis are the following:
\begin{itemize}
\item A non-associated foreground or background eclipsing binary system is blended within the photometric aperture of the target star (BEB).
\item The target is a hierarchical triple system in which two of the components eclipse (HEB).
\item The target star is an eclipsing binary (EB). 
\item A non-associated blended foreground or background star happens to have a transiting planet (Bpl).
\end{itemize}
This list is more comprehensive than MJ11, which only considered BEBs and HEBs quantitatively, and comprises every conceivable astrophysical false positive scenario.  

\section{Procedure}
\label{sec:procedure}

The validation procedure I present in this paper has the following four steps:
\begin{enumerate}
\item Simulate a {\bf representative population} for each scenario, fixing the period to be that of the transit candidate signal.
\item Use this population to calculate the prior for each scenario, optionally taking into account any follow-up observations that may exist.
\item Use the same simulated population to calculate the likelihood of the observed transit signal under each scenario.
\item Combine these numbers to calculate the FPP (Eq. \ref{eq:FPP}) of the signal under an assumption of the {\bf specific planet occurrence rate}; if this number is significantly $<$1\% for a conservative estimate of planet occurrence, then consider the planet validated.
\end{enumerate}
In the following subsections I explain each of these steps in detail, including the important terms in boldface above.

\subsection{Population Simulations}
\label{sec:populations}

Central to the validation procedure I present in this paper is the idea of simulating ``representative populations.''  This means using physically and observationally informed assumptions to generate a population of different instances of a particular transit or eclipse scenario, at a given fixed period, such that the simulated population accurately reflects the true population.  In other words, a representative population must be created such that each instance in the population is equally likely to exist; this makes averaging over the population trivial, as each instance has equal weight in defining the distribution of any desired quantity.  For clarification, in this paper the following boldfaced words conform to this usage: a {\bf representative population} for a particular {\bf scenario} (e.g.~BEB) is made up of many simulated {\bf instances} of that scenario.  For all the results presented in this paper, I create each representative population to have 20,000 instances.

These simulations necessarily require several assumptions:
\begin{itemize}

\item {\bf The population of stars in a cone around the line of sight.}  This requires assumptions about both Galactic structure and stellar evolution.  TRILEGAL (TRIdimensional modeL of thE GALaxy; Girardi et al. 2005, and used in MJ11) is a tool perfectly suited to this purpose.  TRILEGAL takes as input a given direction and area on the sky, a value for extinction at infinity, and parameterizations of the Galactic structure (e.g.~scale height of the disk), and returns an appropriately simulated stellar population, with apparent magnitudes in any chosen photometric system as well as the physical properties of all the stars.

\item {\bf The properties of multiple star systems.}  This encompasses both multiplicity fractions and the distributions of physical and orbital properties for multiple systems, such as the mass ratio and eccentricity distributions.  All these properties may be adopted based on the statistics of observational multiplicity surveys, such as \citet{raghavan2010}.  To simulate the physical distributions of multiple systems, I first start with a total mass of a star system, split it into two components with 50\% probability, and then split one of those into two more with 25\% probability.  Each time I make a split I choose a mass ratio uniformly between 0.1 and 1.  There is no specific physical motivation for this procedure, but it reproduces well the observations of \citet{raghavan2010}, who observed roughly 35-40\% of stars to be in binary systems and about 12\% to be in triple or higher-order multiple systems, and a flat mass ratio distribution for binary pairs.  Besides the binary fraction, I also must assume a ``close binary fraction,'' which is the fraction of stars that have companions with periods small enough to mimic the observed planet transit candidate population.  For this cutoff I choose $P < 300$ days; according to the period distribution observed by \citet{raghavan2010} this encompasses about 12\% of binary stars.  Therefore I assume a ``close binary fraction'' of $0.05 \approx  0.4 \times 0.12$.  Unlike the binary fraction, we do not use a pared-down ``close hierarchical fraction,'' under the assumption that most systems in triple and higher-order systems will undergo significant orbital evolution over their lifetimes, ending up with at least one close pair---and according to studies that have observed this \citep{tokovinin2006}.


\item {\bf Stellar models.}  Accurately simulating eclipsing star systems requires assigning appropriate physical and observational properties to simulated binary companions.  Given a primary star of a particular mass, age, and metallicity, 
a simulated binary companion is assigned a mass according to the assumed mass ratio distribution, and then a radius and magnitudes in different passbands according to stellar model predictions for that particular mass, age, and metallicity.  Models are also required to accurately simulate a target star as a hierarchical system, which requires estimating absolute magnitudes for the target star in addition to its companions, in order to calculate both blended eclipse depths and the total colors of the system.  In this work I use the Padova stellar models \citep{girardi2002} for all masses $>$0.15 $M_\odot$ (the lower limit returned by the online interface that provides the models) and the Dartmouth models \citep{dotter2008} below 0.15 $M_\odot$.  Preference is given to the Padova models for the sake of consistency, because they are used to populate the TRILEGAL simulations.  Below 0.11 $M_\odot$ (the lower limit for the Dartmouth models), I use the models of \citet{baraffe2002}.


\item {\bf Planet occurrence rate.}  Somewhat paradoxically, in order to complete the FPP calculation for any given transit signal, prior assumptions about the planet occurrence rate and radius distribution must be made.  See section 3.4 for how I treat this delicate issue by introducing the concept of the {\bf specific occurrence rate}.  For the background transiting planet scenario, I use a generic assumption of a 40\% planet occurrence rate, with a power law distribution of planet radii $dN/dR \propto R^{-2}$.

\end{itemize}

Due to all these assumptions, the procedure I present in this paper is model-dependent.  However, this is a feature of all Bayesian analysis, which always depends on assumptions (priors) and models, whether physical or empirical.  The advantage of the Bayesian approach is that it provides a natural way to incorporate the existing body of knowledge into a new analysis, and to make assumptions explicit.  That is, there has been significant work in astronomy dedicated to investigating Galactic structure, stellar models, and stellar multiplicity, and the strategy of this analysis is to attack this probabilistic validation problem armed with as much of this prior knowledge as possible.  

\subsection{Priors}
\label{sec:priors}

I define the prior ($\pi_i$) for each scenario to be the product of three distinct factors: the probability that the astrophysical scenario in question exists within the photometric aperture, the geometric probability that the orbit is aligned such that an eclipse is visible, and the probability that the eclipse is ``appropriate''---that is, able to mimic (or be detected as) a transiting planet.  I calculate the first of these factors based on the Galactic population and stellar multiplicity rate assumptions discussed in \S\ref{sec:populations}, combined with knowledge of the sky area inside which a FP might reside.   

The probability of an eclipse occurring, given random orbital orientations with respect to the plane of the sky, is 
\begin{equation}
{\rm Pr}_{\rm ecl} =  \left(\frac{R_1 + R_2}{a}\right) \left(\frac{1 \pm e \sin \omega}{1-e^2}\right),
\end{equation}
where $R_1$ and $R_2$ are the radii of the two bodies in question; $a,e$ and $\omega$ are their orbital semimajor axis, eccentricity, and argument of periastron; and the $\pm$ symbol is a $+$ for primary eclipse and $-$ for secondary.\footnote{Note that using the sum of the radii in this equation for transit probability accounts for the possibility of grazing eclipses; this is different from MJ11, who used the difference of the radii, explicitly rejecting grazing eclipses as potential false positives, on the grounds that any explicitly V-shaped transit could be vetted as a probable false positive based on photometry alone.  In this work, I include the possibility of grazing transits in all the scenarios.}  Calculating this probability for a particular scenario requires averaging over the distribution of the various physical and orbital quantities of that particular scenario (i.e. larger stars are disproportionately likely to host a binary star that eclipses, for a given period); in practice this is accomplished simply by simulating a representative population (\S\ref{sec:populations}) for a particular scenario, assigning isotropic orbital inclinations, and counting what fraction result in an eclipse (either primary, secondary, or both).   

In order for an eclipse of a certain scenario to be ``appropriate,'' it must pass all vetting constraints available, and the probability that a certain scenario will result in an appropriate eclipse may be calculated by just counting what fraction of a representative population pass the constraints.  For the analysis in MJ11, the only constraint we used was depth: the diluted primary eclipse depth must be deep enough to be detected, while the secondary eclipse depth must be shallow enough {\it not} to be detected, given a detection threshold (or one of the eclipses must be missing due to an eccentric orbit).  In the present work, I have generalized the framework in order to naturally include different types of supplementary observations---either preparatory or follow-up, as the case may be.

\begin{figure}[htbp]
   \centering
   \includegraphics[width=3.5in]{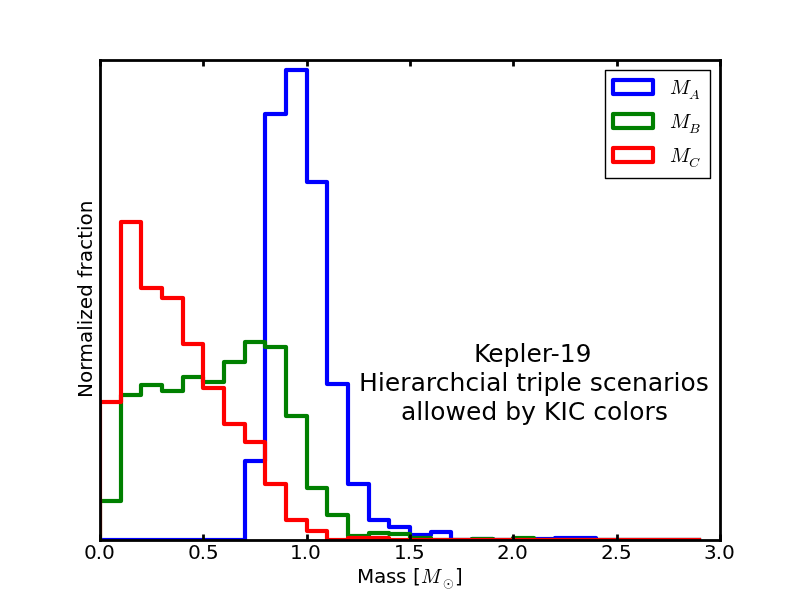} 
   \caption{The distribution of hierarchical triple systems allowed for Kepler-19b (KOI 84.01) by the KIC $g-r$ and 2MASS $J-K$ colors.  Note how there is a tail of higher-mass systems that can mimic Solar-like colors.  This is important for calculating FPP, because systems with higher-mass (and thus higher-luminosity) primaries are more prone to cause false positives with shapes similar to planet transits.  If the spectroscopic properties of the primary are measured (e.g.~with a high-resolution spectrum), then this significantly constrains the false positive landscape (see Figure \ref{fig:heb_spec19b}).}
   
   \label{fig:mhist19b}
\end{figure}

For example, as part of the simulations, I assign each false positive instance apparent magnitudes in many passbands.
This allows observed colors of the target star to be used to constrain the population simulations.  For example, in the case of {\Kepler} targets, each star has {\Kepler} Input Catalog (KIC) $griz$ + 2MASS $JHK$ magnitudes, and we constrain our representative populations of HEB and EB scenarios to conform to the observed total $g-r$ and $J-K$ colors, to within 0.1 mag (I choose these two colors in particular, and not $g-K$, for example, to avoid being too sensitive to metallicity effects \citep{johnsonapps2009}).  This results in asymmetric distributions of target star properties, as shown in Figure \ref{fig:mhist19b}.  If a high-resolution spectrum is obtained for the target star, allowing for spectroscopic measurements of $T_{\rm eff}$ and $\log g$ of the star that dominates the system luminosity, then these populations may be constrained to require the primary star to have the measured properties.

In addition, for each blended false positive instance I assign an angular separation from the target star on the sky, assigned differently for different scenarios.  An important quantity for this step is the so-called ``confusion radius,'' or the separation from the target star inside which a blend might reside.  This may be constrained either by centroid analysis or simply by the size of the aperture, to be most conservative.  Chance-aligned blends are assigned positions perfectly uniformly within the confusion radius surrounding the target star, whereas the positions of hierarchical scenarios are assigned according to the assumed orbital distributions and random mean anomalies.  

Using these separations in combination with the multi-band photometry information, I may use a contrast curve (magnitude contrast as a function of radius from the target star) that results from analyzing a high-resolution  follow-up image in any passband to rule out all instances of a particular scenario that {\it could have been detected} by the observation.  The same principle may be naturally used to rule out instances based on other types of follow-up, such as multi-color transit observations testing for color-dependent transit depth.
After including all available supplementary observational information---or none, as the case may be, in which case the only constraints are the primary/secondary depth conditions and the confusion radius---the ``appropriate probability'' is then the fraction of instances of a particular scenario that pass all the vetting tests.  

\subsection{Likelihoods}
\label{sec:likelihoods}

	Once the representative population for a given scenario is simulated and vetted according to whatever information is available, then the likelihood $\mathcal L$ for a given transit signal under that scenario may be calculated.  It is at this stage in the procedure that the method introduced in this paper advances most significantly beyond the preliminary work of MJ11, so I will first give a brief review of how we treated likelihoods in that work.
	
In a Bayesian calculation, the likelihood factor may be colloquially described as ``the probability of the data given the model.''  In the simplest case, if the data consists of a single measurement, and the property being measured follows a known probability distribution, then the likelihood of the data is simply the known probability density function (PDF) of the property evaluated at the location of the measurement.  This is exactly how we treated the likelihood factor in MJ11: the ``datum'' was the depth of the transit signal, and the PDF was determined using the distribution of depths from the representative population simulations.  
	
The crucial step forward I take in this work is to take full advantage of fact that there is much more information in the shape of a transit signal than just its depth.  In short, transiting planet and diluted eclipsing binary signals, at a fixed period, generally do not look the same.  This is the starting point and backbone of the {\it Kepler} team's BLENDER analysis, in which the goodnesses-of-fit of exhaustive simulations of putative false positive light curve models are compared to the goodness-of-fit of a transiting planet model; false positive models that do not produce acceptable fits are ruled out.
	
In contrast to BLENDER, I do not fit physical models to light curves to rule out false positive scenarios.  Instead, I use the transit signal shape by extending the likelihood analysis of MJ11 into two additional dimensions: duration and slope of the signal, in addition to depth.  I accomplish this by using our representative population simulations to define the three-dimensional probability distribution of these transit shape parameters, and then evaluating this PDF for the shape of the observed signal---the exact three-dimensional analogue of the likelihood calculation of MJ11.  To enable this, I turn to the simplest of transit shape models: the trapezoid.  
	
For each instance in the representative population of each scenario, I follow a two-step procedure.  First, I model the exact shape of the eclipse (accounting for every detail of the eclipsing system including the eccentricity and orbital orientation provided by the simulation) using a limb-darkened eclipse model \citep{MA02} with appropriate quadratic coefficients \citep{claret2000}, taking into account the 30 minute {\it Kepler} long-cadence integration time.  Secondly, I fit that physically modeled light curve with a simple trapezoid function.  I parameterize the trapezoidal shape with three parameters: depth, total duration (from first to last contact), and the ratio of total duration to ingress time, or ``slope,'' which parametrizes the shape of the signal: i.e.~whether it is ``box-shaped,'' ``V-shaped,'' or somewhere in between.  From each population I thus obtain a three-dimensional Òscatter plotÓ of shape parameters, which I turn into a three-dimensional PDF using Gaussian kernel density estimation (using the `gauss\_kde' routine from the scipy.stats Python library\footnote{http://docs.scipy.org/doc/scipy/\\reference/generated/scipy.stats.gaussian\_kde.html}).  
	
Once this PDF has been created for each scenario, the each likelihood may be calculated as well.  To this end, I fit the trapezoidal shape to the observed phase-folded transit light curve, using a Markov Chain Monte Carlo routine \citep{emcee} in order to obtain reliable error bars (important for signals with lower signal-to-noise ratios).  From the MCMC chains I construct a three-dimensional posterior PDF, and then calculate the likelihood by multiplying the two PDFs (``model'' and ``data'') together and integrating over all three dimensions.  This is just like the simple likelihood calculation discussed above, except that instead of evaluating the model PDF at a single point, I integrate over the uncertainty in the data values.  The procedure is illustrated in the figures in \S\ref{sec:tests}, which also give a sense of what the 3D shape distributions look like for different scenarios.  

Note that nowhere in this process do I attempt to derive any physical properties (e.g.~impact parameter, $a/R_\star$ etc.) from the trapezoidal fit, nor do I at any point fit a physical transit model to the observed data---the trapezoid function is purely descriptive, and precise physical properties of the transiting planet are unnecessary in order to calculate its FPP.  I also do not use the candidate's period in the likelihood analysis, apart from the fact that all the simulated scenarios are simulated to have the same period as the candidate.  In other words, I do not attempt to compare the period distributions of the various false positive scenarios with an assumption regarding the planet period distribution. The only question I aim to answer with this likelihood calculation is ``How much does this signal look like each scenario?'' and ultimately ``Does this signal look much more like a transiting planet than a false positive?''  


\subsection{Final Calculation}
\label{sec:finalcalculation}

Besides the priors and likelihoods for all the false positive scenarios, calculating the FPP for a transit signal requires assumptions about the true distribution of transiting planets.  Here my strategy departs from MJ11.  In MJ11, we calculated FPPs under the assumption of an overall 20\% planet occurrence rate, and a distribution of planet radii $dN/dR \sim R^{-2}$.  In order to avoid such blanket assumptions of planet occurrence rates and radius distributions in this work, I proceed as follows.
	
Rather than use a generic distribution of planet radii in order to simulate the representative population of transiting planets that I need (in order to determine the TP prior and likelihood necessary to complete the FPP calculation), I instead simulate a distinct ``radius bin'' population, customized to the signal in question. In other words, I draw planet radii from a uniform distribution between 2/3 and 4/3 the best-fitting planet radius and simulate the representative population of transit signals using these radii.  Of course, this nearly always makes the depth of the signal lie very near the middle of the depth distribution;  this is by design.  If the duration and slope parameters also lie near the middle of the TP distribution while being on the outskirts of all the FP distributions, this may by itself be evidence enough to validate the planet, as I demonstrate in \S\ref{sec:tests}.   
	
\begin{figure}[htbp]
   \centering
   \includegraphics[width=3.5in]{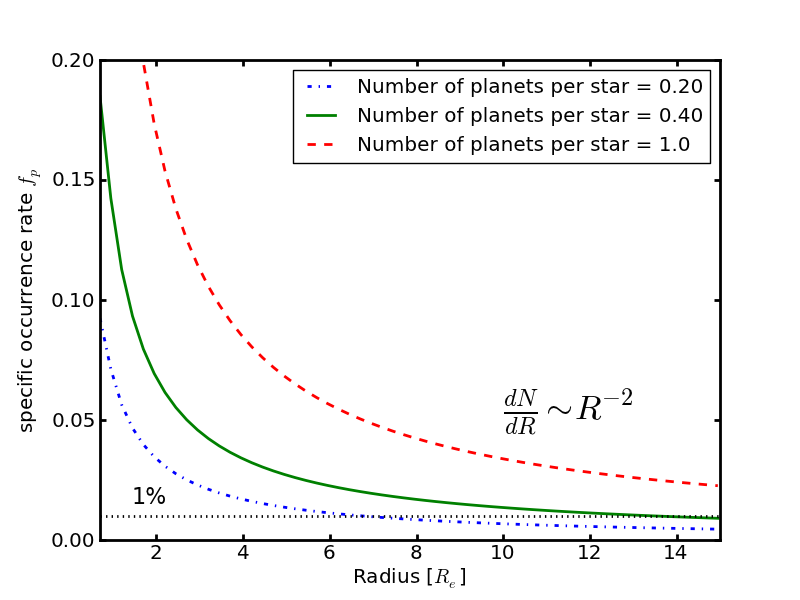} 
   \caption{The specific occurrence rate $f_p$ as a function of radius, for three examples of generic radius distributions, all scaling as the inverse square of the planet radius. I define the specific occurrence rate at a particular radius $R_p$ to be the integral of the planet radius probability density function between $0.7 R_p$ and $1.3 R_p$ (Equation \ref{eq:fp}).  Even though the use of the specific occurrence rate in the analysis presented in this paper makes it independent of assumptions of the overall planet rate or the exact shape of the radius distribution, this illustrates what reasonable values of $f_p$ might be.  For example, it is clear that $f_p = 0.01$ is quite conservative for nearly any planet size, especially for small planets.}
   \label{fig:fp}
\end{figure}

While simulating the planet population to have just the right size to match the observed signal may seem to give an improper advantage to the planet model, this is perfectly reasonable as long as the prior is treated appropriately.  No longer may a generic planet occurrence rate (e.g.~``40\% of stars have planets'') be used; instead, the relevant occurrence rate is the assumed fraction of stars that have planets {\it in the radius bin in question}.  I call this the {\bf specific occurrence rate} $f_p$ and is defined as an integral over the planet radius distribution function $\Phi_R$:
\begin{equation}
\label{eq:fp}
f_p(R_p) = \int_{0.7 R_p}^{1.3 R_p} \Phi_R(R) dR 
\end{equation}
What sort of assumption is reasonable for this $f_p$?  For example, if $\Phi_R(R) \propto R^{-2}$ from 0.5 and 20 Earth radii, 34\% of the probability lies between 0.7 and 1.3 $R_\oplus$, whereas only 3.4\% of the probability lies between 7 and 13 $R_\oplus$.   Thus if an overall 40\% occurrence rate were assumed with this radius distribution, the specific occurrence rate of a $\sim$1 $R_\oplus$ planet should be $\sim$14\% and the specific occurrence rate of a ~10 $R_\oplus$ planet should be 1.4\% (see Figure \ref{fig:fp} for more illustration). 

However, the motivation for this ``radius bin'' strategy is to avoid relying on an assumption of an overall planet occurrence rate and radius distribution, and integrating a generic power law to obtain a specific occurrence rate does not accomplish this.  Instead, consider the following rearrangement and simplification of Equation \ref{eq:simplebayesLs}:
\begin{equation}
\label{eq:fppagain}
{\rm FPP} = \frac{L_{FP}}{L_{FP} + f_p L_{TP}},
\end{equation}
where 
\begin{equation}
L_{FP} = \sum_i \pi_i \mathcal L_i,
\end{equation}
with $i$ representing all of the false positive scenarios, $f_p$ being the true specific occurrence rate, and $L_{TP}$ equalling $\pi_{TP} \mathcal L_{TP}$ if the specific occurrence were unity (i.e.~assuming every star has a planet in this radius bin).  Equation \ref{eq:fppagain} may be further simplified as
\begin{equation}
\label{eq:FPPsimplified}
{\rm FPP} = \frac{1}{1 + f_p P},
\end{equation}
where $P = L_{TP} / L_{FP}$.  This allows me to parametrize the FPP in terms of the unknown specific occurrence rate $f_p$ and the factor $P$, which is independent of assumptions about either planet occurrence or the true planet radius distribution.  

The final result of this validation method may thus be presented as simply this factor $P$, which contains all the information necessary to calculate the FPP based on an assumed specific occurrence rate (Equation \ref{eq:FPPsimplified}).  This lends ease to interpretation, as for large values of $P$, the FPP is very nearly $1 / (f_p P)$.  Thus if $P=1000$, then under an assumption of $f_p = 0.5$, the FPP is 1 in 500.  For $f_p = 0.2$, FPP would be 1 in 200, and so on.  


Another useful presentation answers the following question: at what value of the assumed specific occurrence rate would the FPP be low enough to consider the planet validated?  This is simply a matter of choosing a target FPP (e.g.~${\rm FPP}_V$ for ``validation FPP'')  and solving for $f_p$ in Equation \ref{eq:FPPsimplified}: 
\begin{equation}
\label{eq:fpv}
f_{p,V} = \frac{1 - {\rm FPP}_V} {P \cdot {\rm FPP}_V}
\end{equation}
which may be approximated by $f_{p,V} = (P \cdot {\rm FPP}_V)^{-1}$.   For example, for a target FPP of 0.5\%, $P = 1000$ requires $f_{p,V}=0.2$ to validate, while $P = 2 \times 10^4$ requires only $f_{p,V}=0.01$.  Recent results from both RV surveys \citep{howard2010} and {\it Kepler} \citep{howard2011} suggest that over 40\% of stars have planets of some sort within $\sim$50 day orbits, and that the frequency of planets rises with decreasing radius.  Other work indicates both that the average number of planets per star may even be $>$1 \citep{youdin2011} and that fraction of stars with planets may also be close to unity \citep{cassan2012}.  Thus, if $f_{p,V}$ for a particular transit signal is measured to be significantly below the curves plotted in Figure \ref{fig:fp} (e.g.~if it is 2-3\% for a signal of a $2 R_\oplus$ transit), the planet may be considered securely validated. 


\section{Tests}
\label{sec:tests}

In this section I present the results of applying this procedure to a sample of {\it Kepler} Objects of Interest (KOIs), all of which have been investigated already by other means.  I show in the following subsections that this analysis can both easily validate known planets (using no or very minimal follow-up observations) and identify probable false positives.  \S\ref{sec:kep19b} illustrates the procedure in detail using the example of Kepler-19b, \S\ref{sec:testsample} describes the results of analyzing 17 additional known {\it Kepler} planets, and \S\ref{sec:knownfps} discusses application to 13 known or suspected false positives.

\begin{figure}
   \centering
   \includegraphics[width=3.5in]{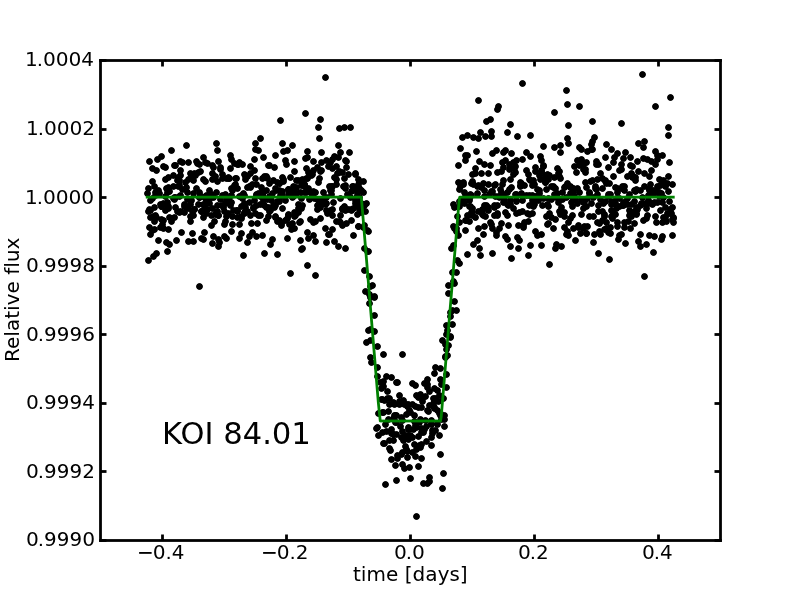} 
   \caption{The phase-folded {\it Kepler} photometry of Kepler-19b (KOI 84.01).  The solid line illustrates the best-fitting trapezoid model used for the likelihood calculations (\S\ref{sec:likelihoods}).  Note that the transit is clearly box-shaped, not V-shaped.}
   \label{fig:signal19b}
\end{figure}

\subsection{Kepler-19b}
\label{sec:kep19b}

\begin{figure*}[ht]

\centering
\subfigure[]{
\includegraphics[scale=0.4]{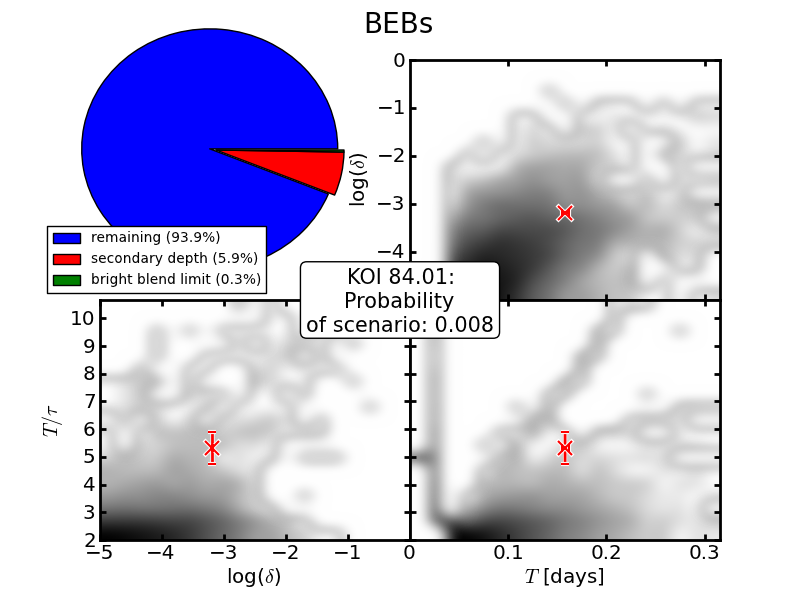}
\label{fig:bgeb19b}
}
\subfigure[]{
\includegraphics[scale=0.4]{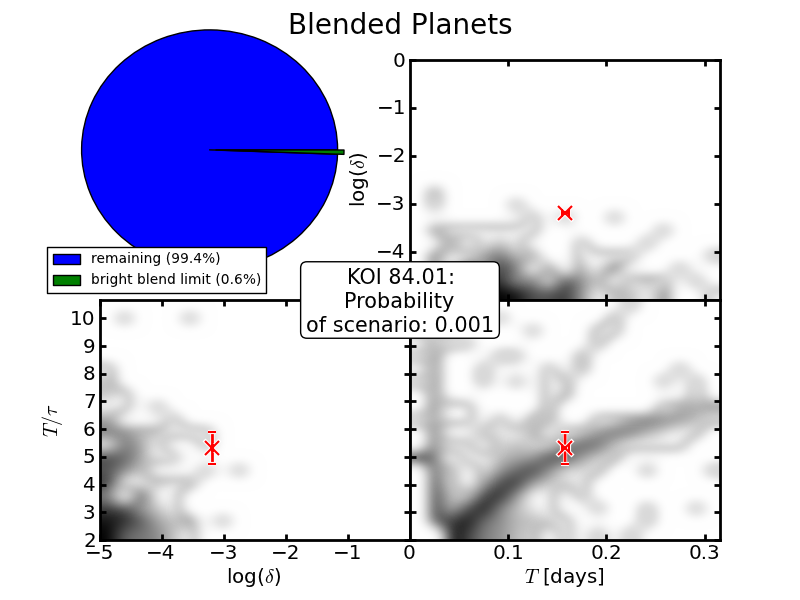}
\label{fig:bgpl19b}
}
\subfigure[]{
\includegraphics[scale=0.4]{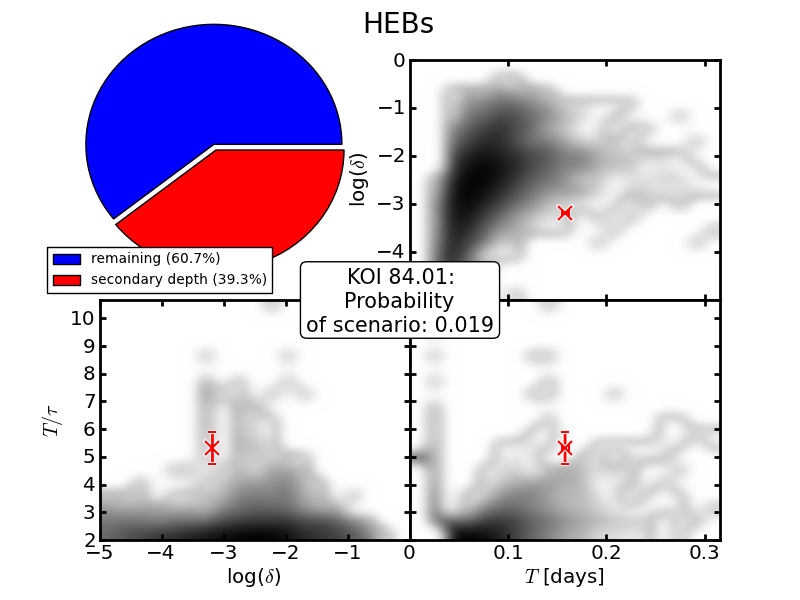}
\label{fig:heb19b}
}
\subfigure[]{
\includegraphics[scale=0.4]{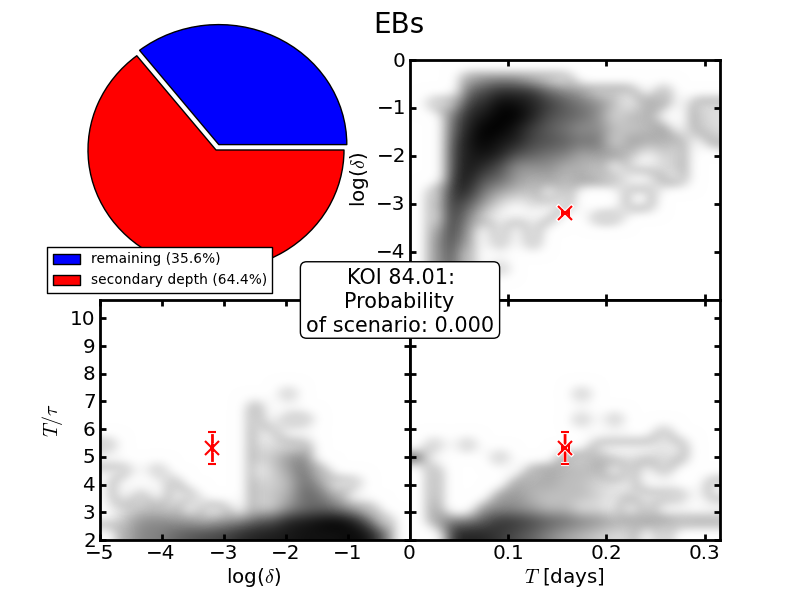}
\label{fig:eb19b}

}

\caption{The false positive landscape for Kepler-19b (KOI 84.01).  Each of these plots illustrates a three-dimensional probability distribution for the trapezoidal shape parameters (depth $\delta$, duration $T$, and ``slope'' $T/\tau$) for a false positive scenario (see \S\ref{sec:framework} for descriptions of the scenarios).  Each of these distributions is made by simulating a statistically representative population (\S\ref{sec:populations}) for a scenario and fitting the shape parameters to each simulated instance of the scenario.  Each population begins with 20,000 simulated instances, and only instances that pass all available observational constraints are included in these distributions---in this case the constraints are that the blended star be at least 1 mag fainter than the primary target, the lack of an observed secondary eclipse deeper than 200ppm, and that the KIC $g-r$ and 2MASS $J-K$ colors match within 0.1 mag.   The pie chart for each scenario illustrates what fraction of the initial simulations pass these tests. On each plot the shape parameters of the transit signal are marked, with the `X' showing the median of an MCMC fit, and error bars illustrating 95\% confidence.  The probability of each scenario is calculated by dividing the prior $\times$ likelihood for that particular scenario by the sum of prior $\times$ likelihood for all the scenarios (including the planet scenario illustrated in Figure \ref{fig:pl19b}).  Priors for each scenario are calculated according to \S\ref{sec:priors}, and the likelihoods are calculated by integrating the illustrated distributions over the observed measurement.
}
\label{fig:fpp19b}
\end{figure*}

\begin{figure}
   \centering
   \includegraphics[width=3.5in]{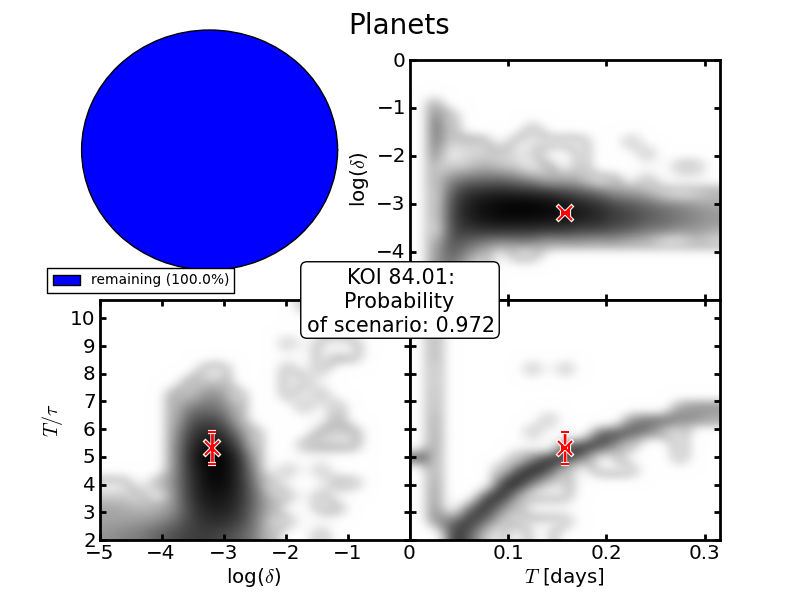} 
   \caption{The transiting planet scenario for Kepler-19b (KOI 84.01).  The plots illustrate the three-dimensional probability distribution for the trapezoidal shape parameters for simulations of a representative population (\S\ref{sec:populations}) of transiting planets around KOI 84.  The radius distribution of planets in this population is chosen to be a narrow range centered on the radius derived for KOI 84.01 assuming it is a {\it bona fide} transiting planet.  The spread and shape of this distribution are caused by different-sized planets in the simulation, different stellar properties (according to uncertainties in stellar radius and mass), and by variations in orbital inclination, which changes duration and slope in a correlated manner (bottom-right panel). The assumed occurrence rate of planets in this radius bin for this calculation is 1\% (see \S\ref{sec:finalcalculation} for more discussion about this ``specific occurrence rate'').  The probability of the Kepler-19b signal being a transiting planet, given the false positive landscape illustrated in Figure \ref{fig:fpp19b} (constrained by only the signal photometry and KIC colors) is about 97\%, giving ${\rm FPP} = 0.03$.  The signal does not fall quite exactly in the middle of the $T/\tau$ distribution because this population simulation uses photometrically estimated physical properties (mass, radius) for the the host star, rather than the more accurate spectroscopically derived properties.  Note that uncertainties like this will tend to decrease the planet likelihood, and thus will contribute to {\it over}estimating the FPP.}
   
   \label{fig:pl19b}
\end{figure}

\begin{figure}
   \centering
   \includegraphics[width=3.5in]{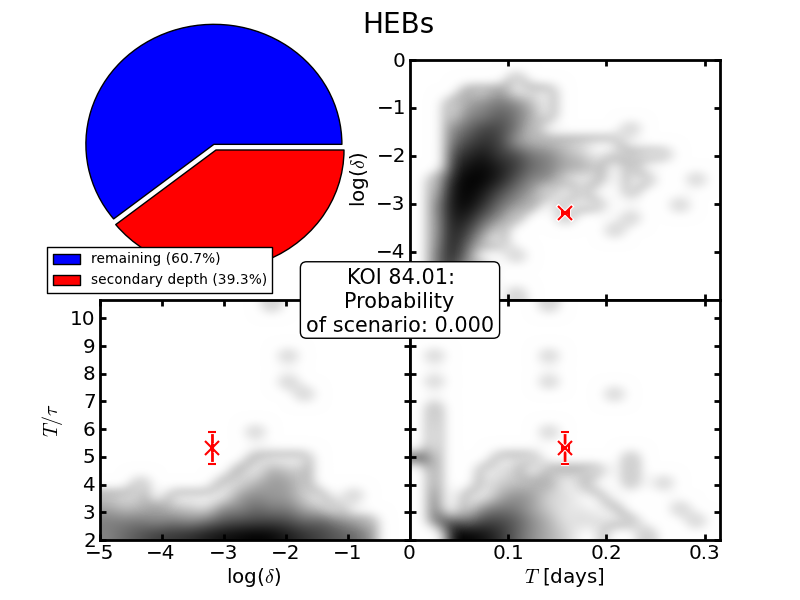} 
   \caption{The hierarchical triple eclipsing binary (HEB) scenario for Kepler-19b (KOI 84.01) under the constraint of spectroscopic characterization of the host star.  The plots illustrate the three-dimensional probability distribution for the trapezoidal shape parameters for simulations of a representative population (¤3.1) of HEB false positive scenarios, but only including those instances in which the spectroscopic properties ($T_{\rm eff}$, $\log g$) of the primary star match those that were measured for KOI 84.  Compare this distribution to Figure \ref{fig:heb19b}:  the systems whose shape parameters overlapped with the properties of the measured signal are not allowed by the spectroscopic characterization---these correspond to the higher-mass systems illustrated in Figure \ref{fig:mhist19b}.  This constraint allows for the signal to be validated, as the HEB scenario was the most-likely FP scenario before including the follow-up spectral information.}
   
   \label{fig:heb_spec19b}
\end{figure}

Kepler-19b, or KOI 84.01, orbits a Sun-like star with a 3.5-day period and has a radius of 2.2 $R_\oplus$.  The discovery paper \citep{ballard2011} details the sequential battery of tests used to rule out false positive explanations for its photometric signal: first the preliminary tests possible with just the photometry (search for a secondary eclipse and measurement of the in-transit photocenter shift), and then the follow-up observations.  The observation follow-up effort included reconnaissance spectroscopy at three epochs using the McDonald Observatory 2.7m telescope, high-precision radial-velocity Keck/HIRES spectroscopy at 20 epochs (though only 8 were used in analysis), adaptive optics (AO) imaging at the Palomar 200-in, speckle imaging at the WIYN telescope in two filters, and 16 hours of {\it Warm Spitzer} observation to measure the color dependence of the transit.  None of these follow-up measurements revealed any sign of a false positive, and BLENDER analysis combined all these constraints with light curve modeling and estimates of the planet and false positive priors (based on {\it Kepler} data itself) to calculate a final false positive probability of about 1 in 7000, validating the signal.

For my analysis of Kepler-19b, I begin from the assumption that no secondary eclipse is detected in the photometry (at a level of 200ppm), and that photocenter analysis constrains any potential blend to be within 2\arcsec at most.  I then proceed through the steps detailed in \S\ref{sec:procedure}: use the target star's position and KIC photometry to simulate the representative populations for each scenario considered, calculate the priors based on these simulations and various assumptions about stellar populations, measure the trapezoidal shape parameters of the Kepler-19b signal itself, and calculate the likelihoods for each scenario based on these measured shape parameters and the distributions of shape parameters of the scenario.  Figures \ref{fig:fpp19b} and \ref{fig:pl19b} illustrate how the shape of the signal matches the typical shape of a transiting planet much better than any of the false positive scenarios.  Assuming a 1\% specific occurrence rate for planets like Kepler-19b, these calculations result in a FPP of 2.8\%.  Alternatively, a specific occurrence rate of about 6\% would be required for FPP=0.5\%--- close to being securely validated.  The most likely false positive is the HEB scenario (Figure \ref{fig:heb19b}).


When generating the representative population for the true transiting planet scenario for this signal, I use the stellar parameters for KOI 84 provided in \citet{batalha2012} (0.86 $R_\odot$, 0.91 $M_\odot$), with 20\% uncertainties.  These notably do not agree exactly with parameters published in the discovery paper, determined from spectroscopic analysis (1.1 $R_\odot$, 0.97 $M_\odot$).  The effect of this discrepancy is visible in Figure \ref{fig:pl19b}, where the measured slope parameter ($T/\tau$) of the signal appears to fall slightly above the main distribution of the simulated population: a smaller star will result in a smaller $T/\tau$ for a fixed value of $R_p/R_\star$.  The result is that the likelihood of the planet signal is calculated to be lower than what it would be if the stellar parameters of the simulated population were correct, leading to an {\it over}estimate of the FPP.  Since accurate stellar parameters will not necessarily always be available, this circumstance makes for a realistic test case.


The next step in my re-analysis of the Kepler-19b FPP is to pretend that a single follow-up observation is available: a high-resolution spectrum from which $T_{\rm eff}$ and $\log g$ of the host star can be accurately measured, to a precision of 80 K and 0.1 dex, respectively.  The effect of this constraint is to rule out those binary and hierarchical scenarios from the population simulations that were initially allowed (since the overall colors matched the KIC colors), but whose temperatures or surface gravities are $>$3$\sigma$ discrepant from this measurement.   Figure \ref{fig:heb_spec19b} shows how the distribution of shape parameters for the HEB scenario changes under this constraint such that the measured Kepler-19b parameters are even more on the outskirts---the HEB instances that could mimic a transit shape were exactly the ones ruled out by the stellar characterization.   Using the likelihoods and priors updated for this constraint, the specific occurrence rate assumption necessary to result in an FPP of 0.5\% ($f_{p,V}$) for Kepler-19b is only 1.8\%---small enough to result in a secure validation.


To investigate whether a single AO observation would have a similar effect, I remove the spectroscopic constraint and submit the simulated populations to the constraint of a generic deep AO image ($\Delta K = 8$ at $0\farcs7$), ruling out all instances the false positive scenarios that would be identified by this observation.  This AO constraint alone also decreases the FPP enough to claim validation, though not quite as strongly as the spectoscopic constraint: $f_{p,V} = 0.03$.  The combination of the two constraints results in $f_{p,V} = 0.1\%$.

The conclusion of this investigation is striking: secure validation of the Kepler-19b signal with these methods would have been possible using either of two single-epoch follow-up observations: a high-resolution spectrum to measure the spectroscopic parameters or an AO image to constrain blends.   With both of these observations, validation would be secure beyond doubt.

\begin{deluxetable}{clcccc}
	\tabletypesize{\footnotesize}
	\tablecolumns{6}
	\tablecaption{Test sample: Non-TTV Confirmed Kepler planets\tablenotemark{1}}
	\tablehead{ \colhead{Name} &\colhead{Method} & \colhead{$f_{p,V}$} &  \colhead{$f_{p,V}^{\rm spec}$} &  \colhead{$f_{p,V}^{\rm AO}$} & \colhead{$f_{p,V}^{\rm spec,AO}$}}
	\startdata
Kepler-4b & RV & \bf{0.006} & \bf{0.013} & \bf{0.003} & \bf{0.010}\\
Kepler-6b & RV & 0.161 & 0.163 & 0.162 & 0.163\\
Kepler-7b & RV & 0.068 & 0.140 & 0.052 & 0.112\\
Kepler-8b & RV & 1.227 & 1.087 & 0.808 & 0.781\\
Kepler-10b & RV/BLENDER & \bf{0.014} & \bf{0.014} & \bf{0.000} & \bf{0.000}\\
Kepler-10c & BLENDER & \bf{0.014} & \bf{0.012} & \bf{0.002} & \bf{0.000}\\
Kepler-11g & BLENDER & 0.289 & 0.200 & 0.140 & 0.061\\
Kepler-12b & RV & 0.086 & 0.018 & 0.086 & 0.018\\
Kepler-14b & RV & 0.010 & \bf{0.000} & \bf{0.000} & \bf{0.000}\\
Kepler-15b & RV & 0.293 & 0.116 & 0.193 & 0.104\\
Kepler-17b & RV & 0.109 & 0.131 & 0.094 & 0.118\\
Kepler-18b & BLENDER & 0.460 & 0.241 & 0.189 & \bf{0.002}\\
Kepler-19b & BLENDER & 0.058 & \bf{0.018} & \bf{0.030} & \bf{0.001}\\
Kepler-20e & BLENDER & 0.071 & 0.056 & \bf{0.014} & \bf{0.001}\\
Kepler-20f & BLENDER & 0.178 & 0.176 & \bf{0.009} & \bf{0.007}\\
Kepler-21b & BLENDER & \bf{0.010} & \bf{0.010} & \bf{0.001} & \bf{0.001}\\
Kepler-22b & BLENDER & 0.048 & 0.051 & \bf{0.004} & \bf{0.001} \\
\hline \\
{\bf TOTALS}  & {\bf 17 (all)} & {\bf 4} & {\bf 6} & {\bf 9} & {\bf 10}
	\enddata
	\tablenotetext{1}{The results of applying this validation procedure to a sample of known {\it Kepler} planets, 9 of which were confirmed with RV measurements and 9 by BLENDER analysis (one by both).  $f_{p,V}$ is the specific planet occurrence rate required to result in an FPP of 0.5\% (Equation \ref{eq:fpv}).  When $f_{p,V}$ is significantly below a reasonable occurrence rate estimate, (Figure \ref{fig:fp}), I consider the planet validated (marked in bold).  The `spec' and `AO' superscripts indicate the effect of incorporating spectroscopic and imaging follow-up observations.  Most of the planets are validated with both types of follow-up; the ones that are not are nearly all giant planets.}
	\label{table:resultspl}
\end{deluxetable}

\subsection{Kepler planet test sample}
\label{sec:testsample}

While Kepler-19b is an eye-opening individual example of the potential of this analysis to fully validate transiting planet candidates with very modest follow-up investment, it is just a single example.  I thus identify a broader sample to test in a similar manner: 17 other known {\it Kepler} planets, selected to be all of the officially designated {\it Kepler} planets that do not show significant transit timing variations. 9 of these were confirmed with RV measurements, and the additional 8 were BLENDER-validated.

For each of these transit signals I follow the exact same analysis sequence as described above for Kepler-19b, using the same generic assumptions of a 200ppm secondary eclipse limit and a blend radius of 2\arcsec.  I find that 4 of the 17 would have been validated with no follow-up observations at all, 6 require only stellar characterization to validate, 9 require only AO imaging, and all except 7 (all giant planets but 1) can be validated with the combination of both constraints.  These results are summarized in Table \ref{table:resultspl}, which presents the values of $f_{p,V}$ for each signal under the various constraints.  A signal is considered validated when $f_{p,V}$ is significantly below the expected planet occurrence rate for a planet of that size---the values low enough for validation are bolded in the table.  These results are quite encouraging, suggesting that perhaps many future {\it Kepler} planet validations will require only minimal follow-up observations.

\begin{figure}
   \centering
   \includegraphics[width=3.5in]{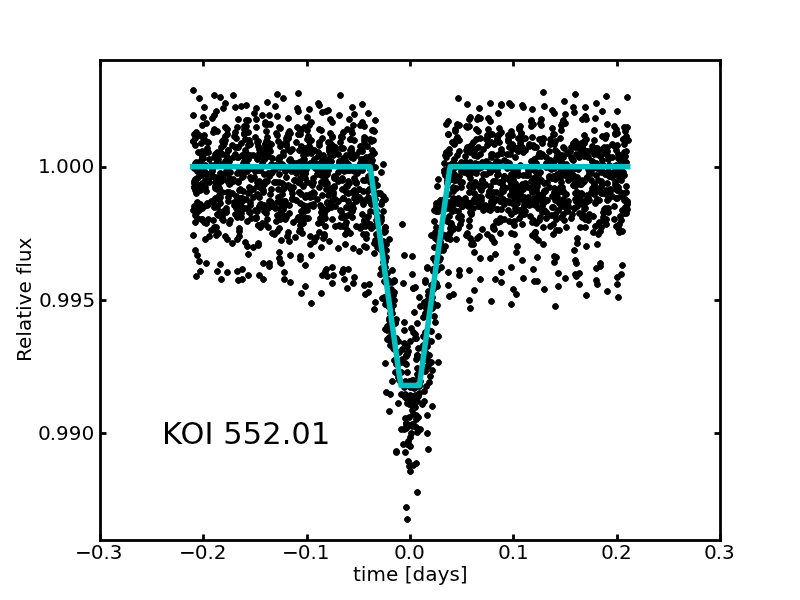} 
   \caption{The phase-folded {\it Kepler} photometry of KOI 552.01.  The solid line illustrates the best-fitting trapezoid model used for the likelihood calculations (\S\ref{sec:likelihoods}).  \citet{bouchy2011} measured large RV variations of the host star suggesting that this signal is caused by an eclipsing M-dwarf rather than a planet.  Note that the signal is clearly more V-shaped than box-shaped.}
   \label{fig:signal552}
\end{figure}

\subsection{Known false positives}	
\label{sec:knownfps}

Despite the encouragement of the above results, it is fair to ask whether this analysis is skewed such that it {\it always} returns a low false positive probability, and thus automatically tuned for easy planet validation.  To address this concern, 
I investigate a sample of {\it Kepler} candidates known to be false positives.

KOI 552.01 (Figure \ref{fig:signal552}) is a Jupiter-sized candidate whose host star has a KIC-estimated $T_{\rm eff} = 6018$ K.  This was selected as one of the candidates for follow-up with the SOPHIE spectrograph, and \citet{bouchy2011} measured it to have $\sim$km/s RV variation between two epochs.  This suggests that the signal is likely caused by an eclipsing M-dwarf star rather than a giant planet.  In addition, analysis of the spectra suggests that the temperature of the star is 6500 K, significantly hotter than the KIC estimate.  

\begin{figure}
   \centering
   \includegraphics[width=3.5in]{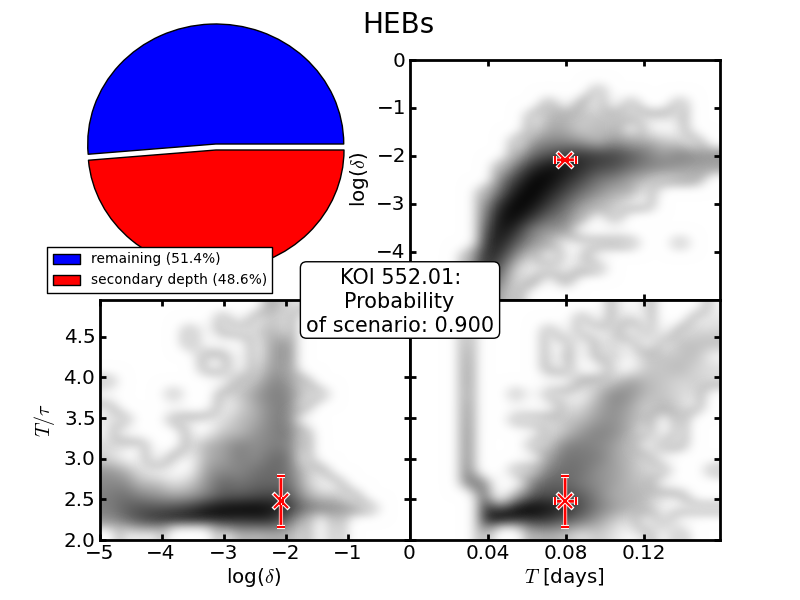} 
   \caption{The hierarchical triple eclipsing binary system scenario for KOI 552.01.  This KOI was measured by \citet{bouchy2011} to have $\sim$km/s radial velocity variation between two epochs, suggesting a stellar eclipse rather than a transiting planet---an observed astrophysical false positive.  The analysis presented in this paper gives an FPP of $>$90\% for this signal, and this figure illustrates why: the shape of the signal is exactly consistent with what would be expected from a hierarchical triple system including an eclipsing binary.}
   
   \label{fig:heb552}
\end{figure}

For KOI 552.01 I calculate FPP=0.90 (assuming a specific occurrence of 1\%, appropriate for Jupiter-sized planets \citep{wright2012})---in other words, this procedure clearly identifies this signal as a very likely false positive using only the photometry, and no follow-up information at all.  Figure \ref{fig:heb552} illustrates how well the shape of the signal is described by a hierarchical system with an eclipsing binary.  Conveniently, an HEB explanation for KOI 552.01 explains not only the eclipse signal but also the discrepancy between the KIC and spectroscopic $T_{\rm eff}$ estimates: if this system is indeed a triple system with a primary with $T_{\rm eff}=6500$ K and two cooler stars, then trying to interpret the broadband photometry of the whole system as a single star would naturally result in a cooler temperature estimate.

Similar analysis of 12 other KOIs that have been identified as false positives gives similar results, with all but one having FPP $>$ 20\%, and six having FPP $>$ 80\% (Table \ref{table:resultsfp}).  Of these, 11 were identified as false positives by \citet{santerne2012} and one (1187.01) was identified by \citet{colon2012} to have a color-dependent transit depth.  I assume a specific occurrence rate of 1\% for all these candidates.  All but 1187.01 have transit depth  $\geq$ 0.4\%; this is due to the way that the SOPHIE follow-up sample was chosen.  Notably, many of these signals have $T/\tau$ parameter very close to 2, indicative of a V-shaped transit shape.  This demonstrates that incorporating transit shape into this analysis allows this procedure not only to validate planets, but also to reliably identify likely false positives. 

\begin{deluxetable}{clccc}
	\tabletypesize{\footnotesize}
	\tablecolumns{5}
	\tablecaption{Test sample: False positives\tablenotemark{1}}
	\tablehead{ \colhead{KOI} &\colhead{FP identification method} & \colhead{$\delta$(\%)\tablenotemark{2}} & \colhead{$T/\tau$\tablenotemark{3}} & \colhead{FPP}\tablenotemark{4}}
	\startdata
552.01 & RV variation\tablenotemark{a} & 0.8 & 2.6 & 0.95 \\
1187.01 & color-dependent transit\tablenotemark{b} & 0.2 & 2.4 & 0.82 \\
190.01 & Blended CCF; RVs\tablenotemark{c} & 1.1 & 3.4 & 0.22 \\
340.01 & RVs\tablenotemark{c} & 2.1 & 5.5 & 0.30 \\
418.01 & Blended CCF; RVs\tablenotemark{c} & 1.2 & 3.4 & 0.85 \\
419.01 & RVs; sec. eclipse only\tablenotemark{c} & 0.8 & 2.4 & 0.96 \\
425.01 & Blended CCF; RVs\tablenotemark{c} & 1.2 & 2.8 & 0.61 \\
607.01 & RVs\tablenotemark{c}  & 0.7 & 3.7 & 0.32 \\
609.01 & double-lined binary\tablenotemark{c} & 0.4 & 2.5 & 0.96 \\
667.01 & blend (DSS photometry)\tablenotemark{c} & 0.9 & 3.8 & 0.07 \\
698.01 & RVs; sec. eclipse only\tablenotemark{c} & 0.7 & 2.5 & 0.96 \\
1786.01 & RVs\tablenotemark{c} & 0.8 & 5.8 & 0.48 

	\enddata
	\tablenotetext{1}{The results of applying this analysis on a sample of KOIs that have been identified as false positives with various follow-up observations.  Nearly all are identified as likely false positives.}
	\tablenotetext{2}{Transit depth}
	\tablenotetext{3}{The ratio of the total transit duration to the ingress/egress duration.  A value close to 2 is a V-shaped signal; a larger value indicates a more box-shaped transit.}
	\tablenotetext{4}{False positive probability, assuming a specific occurrence rate of 1\%.}
	\tablenotetext{a}{\citet{bouchy2011}}
	\tablenotetext{b}{\citet{colon2012}}
	\tablenotetext{c}{\citet{santerne2012}}
	\label{table:resultsfp}
\end{deluxetable}

\subsection{Testing assumptions and simulation variance}

\begin{figure}
   \centering
   \includegraphics[width=3.5in]{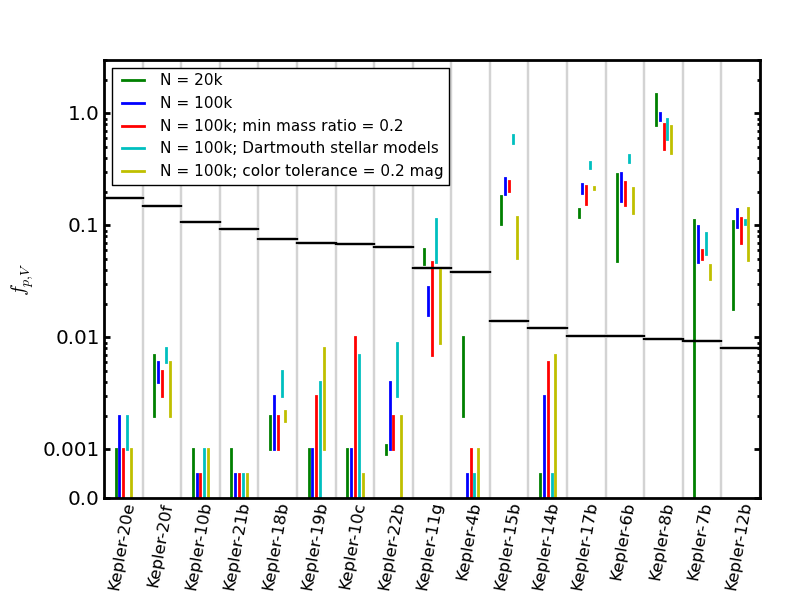} 
   \caption{The results of testing the effects of sample variance and changing various assumptions on the false positive calculations described in this paper.  Each colored vertical line represents the full range of $f_{p,V}$ (the specific occurrence rate assumption required for validation) over three iterations of the procedure for each planet from Table \ref{table:resultspl}, ordered by increasing radius.  The five different vertical for each planet represent 5 different experiments, varying the size of the Monte Carlo simulations and different input astrophysical assumptions.  All of the planets that were claimed as validated by the original analysis (the final column of Table \ref{table:resultspl}) remain validated for each iteration of each of these tests.}
   
   \label{fig:FPPtests}
\end{figure}

As mentioned in \S\ref{sec:populations}, the false positive calculations that this procedure delivers are dependent on a set of astrophysical assumptions.  In addition, apart from \textit{a priori} assumptions, the calculations are subject to intrinsic variance due to the fact that FP distributions used to calculated the likelihoods are the result of Monte Carlo simulations.   To test the degree to which qualitative results depend on these effects, we repeat the calculations for the known planet sample many times, using different sets of assumptions.   I repeat each test three times, with the full range of results for the final $f_{p,V}$ numbers illustrated for each test in Figure \ref{fig:FPPtests}.  The tests are the following:
\begin{itemize}
\item The original simulations under the original assumptions (repeated three times).
\item The original simulations under the original assumptions, but using $N=100,000$ for the simulations (rather than $N=20,000$)
\item $N=100,000$, changing the mass ratio distribution to be flat between 0.2 and 1 (rather than between 0.1 and 1)
\item $N=100,000$, using the Dartmouth stellar models \citep{dotter2008} in place of the Padova models.
\item $N=100,000$, increasing the ``color tolerance'' of the EB and HEB scenarios to 0.2; i.e.~allowing all FP scenarios that match within 0.2 mag of the $g-r$ and $J-K$ colors, rather than 0.1 mag.
\end{itemize}

As seen in the figure, in only two cases do the range of results cross the nominal $f_{p,V}$ ``validation threshold'': Kepler-11g occasionally drops into validation territory, as does Kepler-7b, for one instance of the $N=20,000$ simulation.  None of the planets validated with both follow-up observations from Table \ref{table:resultspl} cross over into the non-validation regime.  We conclude that larger simulations are always better if possible, to reduce variance, but that $N=20,000$ is usually sufficient.  While the exact value of FPP may change by a factor of a few from repetition to repetition, secure validations remain secure validations.

There are other input assumptions not tested by this exercise, but these are mostly trivial assumptions that can be tested on a case-by-case basis simply by adjusting the multiplicative factors that go into the prior calculation for a given scenario.  These include stellar binary and triple fractions, the fraction of binaries that are in ``short'' orbits, the occurrence rate of planets around background stars, and the density of background stars returned by TRILEGAL.  Adjusting any of these numbers changes the prior factor for that scenario, and correspondingly changes the false positive probability.  If the FPP is small and one false positive scenario dominates, the overall FPP is directly proportional to the prior factor for that scenario (see the discussion in \S5 of MJ11); e.g., if the HEB scenario is the most likely false positive for a particular candidate, doubling the assumed stellar triple fraction will approximately double the FPP (or, equivalently, double the $f_{p,V}$ required for validation). 

\section{Relation to Morton \& Johnson (2011)}
\label{sec:MJ11}

While I have demonstrated that the results from this newly updated FPP analysis are reliable both with known planets and known false positives, I now briefly discuss why these results may differ from the individual FPPs published in \citet{mj11}.  First of all, it is important to understand that the goal of the previous work was to demonstrate that, in general, {\it Kepler} candidates {\it that pass all possible initial photometric vetting} are very likely to be planets.  That is, the MJ11 analysis is only strictly valid for signals that are not obviously V-shaped and whose light curves have been declared free from secondary eclipses.  

As it turns out, many of the candidates from \citet{borucki2011} were not in fact fully vetted.  This was made clear in that paper, but this fact was not reflected in the \citet{mj11} KOI FPP table.  In other words, there are some signals that are clearly V-shaped (a number of these are now in the known false positive sample discussed in \S\ref{sec:knownfps}), and likely also signals that show faintly detectable secondary eclipses.  For such KOIs, it is probable that the final FPP will turn out to be significantly higher, as is the case for the sample of false positives detected by \citet{santerne2012}.

The analysis of \citet{mj11} was neither a planet validation procedure nor a false positive detection procedure.  Nor was its purpose to claim that the {\it Kepler} results are completely immune from the problem of false positives.  Instead, its goal was to demonstrate that the {\it a priori} FPP for fully vetted {\it Kepler} candidates is typically low, thus suggesting that the overall sample of candidates generally reflects the properties of the true sample of planets (barring the existence of a pathologically large number of V-shaped signals among the candidate list).  On the other hand, this work, having advanced the techniques of \citet{mj11}, is intended to validate candidates, using all available information on an individual basis.

\section{Conclusions}
\label{sec:discussion}

Transiting exoplanet surveys continue to grow much faster than the capabilities of traditional follow-up infrastructure.  While the observations required to positively confirm candidates have always been the bottleneck for transit surveys, the {\it Kepler} mission has made this issue much more acute.  Future ground- and space-based surveys will doubtless continue to produce orders of magnitude more candidates than can be securely confirmed either with radial velocity or transit timing variation measurements.  As a result, the onus of confirmation for the majority of transiting planet candidates has shifted irrevocably from dynamical confirmation to probabilistic validation.


With probabilistic validation playing such a central role, the procedure I describe in this paper enables a revolution in how transit surveys could operate: {\it validating more planets, using fewer follow-up resources}.  Recall what I have demonstrated about this procedure:
\begin{itemize}
\item It is efficient: an end-to-end FPP calculation for a transit signal takes only about 10 minutes as currently implemented---and there is room for further computational optimization.
\item It often requires only one or two single-epoch follow-up observations (either spectral characterization, AO imaging, or both---and sometimes neither) to validate a true transiting planet.
\item It can reliably identify false positives.
\end{itemize}
The benefits of applying this analysis to a large sample of candidates from a transit survey such as {\it Kepler} are evident.  
Spectroscopic follow-up can be prioritized in order of increasing FPP, likely leading directly to many validations.  Preference for AO imaging can then be given to those candidates not validated with a spectrum.  RV observations of likely false positives can be avoided, and more time devoted to RV measurements of likely planets or characterization spectroscopy.  And the most resource-intensive follow-up tools, such as multi-band photometry, validation-inspired RV measurements, and the full BLENDER analysis, can be reserved for the most stubborn candidates---those that are not likely false positives, but resist validation even after spectral characterization and AO imaging.  A good example of a candidate ideal for this last-stage analysis would be one with a residual FPP of 5-10\% or so, for which AO imaging reveals a faint blended companion.  

With regard to {\it Kepler}, I intend to apply this procedure to {\it all} of the remaining unvalidated KOIs, using whatever follow-up observations are available.  There are several ongoing projects to spectroscopically characterize and obtain AO images of large numbers of KOIs; doubtless these efforts have already provided the measurements necessary to validate hundreds of planets.

The ultimate goal of this project is to make self-contained, stand-alone transiting planet validation software available to the community.  This will allow present and future transit surveys to apply this analysis at an early stage, hopefully enabling  follow-up resources to be optimized and planet validations to be expedited.  The hope is that with fewer observational resources dedicated to the task of planet validation, more can be devoted to detailed characterization of individual systems of particular interest.  In addition, with large numbers of securely validated transiting planets, statistical population studies will become possible without worrying about any residual effects from false positive contamination.  

\acknowledgments{TDM acknowledges helpful discussions with John Johnson, Erik Petigura, Jason Eastman, Justin Crepp, Jon Swift, John Carpenter, Steve Bryson, and many {\it Kepler} team members throughout the course of this research.  He also thanks the referee Scott Gaudi, who waived anonymity, and whose suggestions led to substantial improvements to this manuscript.}

\bibliographystyle{apj}
\bibliography{allrefs}

\end{document}